\documentclass[sigplan,review,nonacm]{acmart}

\AtBeginDocument{%
  \providecommand\BibTeX{{%
    Bib\TeX}}}

\usepackage{amsmath,amsfonts}
\usepackage{graphicx}
\usepackage{textcomp}
\usepackage{xcolor}
\usepackage[utf8]{inputenc}
\usepackage{comment}
\usepackage[]{hyperref}
\usepackage{color}
\usepackage[dvipsnames]{xcolor}
\usepackage{todonotes}
\usepackage{array}
\usepackage{multirow}
\usepackage[noend]{algpseudocode}
\usepackage[export]{adjustbox}
\usepackage{threeparttable}
\usepackage{makecell}
\usepackage{subfig}
\usepackage{diagbox}
\usepackage{tabularx,colortbl,xcolor}

\usepackage{tikz}       
\usepackage{xcolor}    
\usepackage{marginnote} 
\usepackage[noend,linesnumbered,ruled,vlined]{algorithm2e}

\usepackage{soul}
\usepackage[authormarkup=none, commandnameprefix=ifneeded]{changes} 
\definecolor{myblue}{RGB}{0, 0, 255}  
\definecolor{myred}{RGB}{255, 0, 0}    
\colorlet{Changes@ColorAdded}{myblue}  
\colorlet{Changes@ColorDeleted}{myred} 
\usepackage{cancel}
\usepackage{gensymb}
\usepackage{wrapfig}
\usepackage{subcaption}

\DeclareUnicodeCharacter{202F}{\,}
\def\BibTeX{{\rm B\kern-.05em{\sc i\kern-.025em b}\kern-.08em
    T\kern-.1667em\lower.7ex\hbox{E}\kern-.125emX}}
\begin{document}

\pdfpagewidth=8.5in
\pdfpageheight=11in

\pagenumbering{arabic}

\title{EPIC: A System Framework for Efficient Egocentric Perception on Embodied AR Glasses}

\author{
Tianhua Xia \enspace Haiyu Wang  \enspace  Jiajing Zheng \enspace  Su Chen \enspace Sai Qian Zhang\\
New York University\\
}


\begin{abstract}
Modern smart AR glasses are evolving into intelligent systems that support foundation model-based assistance through continuous perception of the user and surrounding environment. However, this perception-first design creates major bottlenecks. Continuously capturing, processing, and storing rich perceptual streams, especially high-resolution egocentric video, imposes substantial power and memory overhead, which is difficult to sustain on resource-constrained AR glasses.
In this work, we propose EPIC, an efficient egocentric perception system for embodied intelligence on smart AR glasses. EPIC is an algorithm-hardware co-optimization framework that leverages gaze, pose, and inertial signals to infer user intent and retain only the most informative parts of high-resolution perceptual input, greatly reducing perception overhead. Our results show that EPIC reduces memory footprint by $27.5\times$ and energy consumption by $24.3\times$ on average compared with full video baseline solution, while preserving intelligent assistance accuracy on egocentric video understanding tasks, a key application scenario for embodied intelligence on smart glasses.
\end{abstract}

\maketitle
\thispagestyle{plain}
\pagestyle{plain}



\section{Introduction}
\label{sec:intro}
Augmented reality (AR) is a transformative technology that is reshaping how humans interact with digital information and the physical world. By overlaying context-aware digital content directly onto a user’s real environment, AR enables more natural, immediate, and intuitive access to information than traditional screen-based interfaces. This capability makes AR valuable across a wide range of applications, including education~\cite{westin2022inclusive, al2023analyzing, takrouri2022ar}, healthcare~\cite{chirico2016virtual, viglialoro2021augmented, hsieh2017vr} and manufacturing~\cite{bottani2019augmented, sahu2021artificial, nee2012augmented} where real-time perception, guidance, and interaction are essential.

\begin{figure}
    \centering
    \includegraphics[width=\columnwidth]{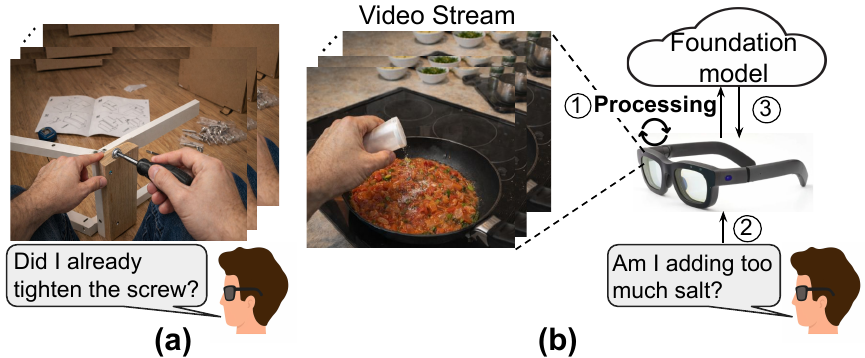}
    \caption{(a) An example on daily assistance on smart AR glass. (b) The detailed system workflow of embodied AI assistance, step numbers are shown in circles.}
    \label{fig:framework}
\end{figure}

Today’s smart AR glasses are still at an early stage, but they are beginning to evolve from simple display and interaction devices into intelligent systems that depend on continuous perception of both the user and the surrounding environment to provide context-aware assistance. This shift represents an important future direction for AR systems. For example, as illustrated in Figure~\ref{fig:framework} (a), when a user wearing AR glasses is assembling furniture and asks, ‘Did I already tighten this screw?’, the device may need to analyze a sequence of recent video frames and send them to AI models, such as embodied foundation models (EFMs)~\cite{zheng2025universal}. In another scenario, when a user is cooking and asks whether the right amount of salt has been added, answering the question may require an EFM to reason over a longer video stream captured throughout the cooking process, as shown in Figure~\ref{fig:framework} (b). More broadly, AR platforms provide a natural foundation for~\textbf{embodied intelligence}~\cite{gupta2021embodied, liu2025embodied}, in which smart glasses continuously leverage perceptual signals, such as egocentric video, to infer user context, reason about ongoing activities, and deliver incremental, context-aware assistance~\cite{fung2025embodied,lampropoulos2025intelligent}.

Figure~\ref{fig:framework} (b) also illustrates the detailed system flow of embodied intelligence~\cite{fung2025embodied,lampropoulos2025intelligent}. 
However, supporting such intelligent assistance on battery-powered AR devices remains challenging. Streaming and buffering perceptual data place heavy pressure on limited memory and storage, while continuous sensing and preprocessing rapidly consume energy. As these applications become more common in daily life, AR systems must repeatedly capture, buffer, and process high-resolution sensor streams, especially video, to preserve enough context for downstream reasoning. Although offloading video to the cloud can reduce local storage demand, it does not remove the high energy cost of continuous capture and transmission, and it also depends heavily on reliable network connectivity. As a result, modern AR systems still need to buffer a significant portion of perceptual data locally, making efficient on-device memory management a key system challenge~\cite{meta_video,magicleap_video,apple_video}.

\begin{figure*}
    \centering
    \includegraphics[width=2\columnwidth]{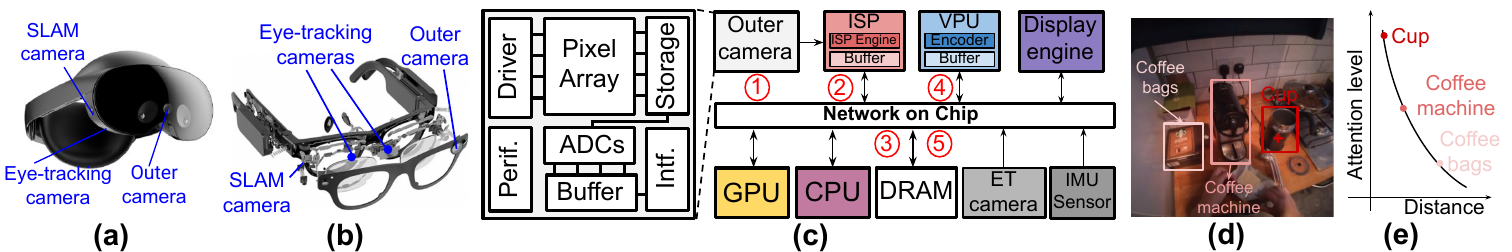}
    \caption{(a) Meta Quest Pro. (b) Meta Orion Smart AI Glass. (c) The SoC layout of the AR device together with the processing flow for video stream, step numbers are shown in circles. (d) An example illustrating how user attention varies with gaze location. (e) Attention variation changes with distance.}
    \label{fig:bg_soc}
\end{figure*}

To mitigate this problem, we adopt an algorithm–hardware co-optimization approach to enable low-power, low-footprint perception for embodied intelligence. To this end, we propose~\textit{EPIC}, an~\underline{E}fficient Egocentric~\underline{P}erception Framework for Embodied~\underline{I}ntelligen\underline{c}e on AR Glasses. EPIC leverages multimodal AR signals to track user motion and intent, enabling it to effectively eliminate redundant information in egocentric video streams across both spatial and temporal dimensions. Our contribution can be summarized as follows:
\begin{itemize}
\item We propose EPIC algorithm, a deep learning based solution that leverages user motion and gaze location to exploit spatial and temporal correlations for efficient video stream compression. To enable fine grained selection of important image patches for storage, EPIC introduces an adaptive patch storage protocol that maximizes redundancy elimination.
\item We propose the EPIC hardware accelerator as a plug-in to AR SoC to significantly reduce the cost of video processing. We also introduce a lightweight enhancement to the image sensor that lowers image transmission cost with minimal hardware changes.
\item  Evaluation results show that EPIC reduces the memory footprint by $27.5\times$ and energy consumption by $24.3\times$ on average compared to full video solution while preserving intelligent assistance accuracy on egocentric video understanding tasks.
\end{itemize}


\section{Background and Related Work}
\label{sec:bg}
\subsection{AR Device Overview}
\label{sec:bg:AR_device}

Modern AR devices integrate a diverse set of sensors that continuously capture both environmental and user centric signals with relatively low overhead. As shown in Figure~\ref{fig:bg_soc} (a) and (b), platforms such as Meta Quest Pro~\cite{meta_quest} and Meta Orion glasses~\cite{meta_orion} include tightly integrated sensing systems for egocentric perception. 
Collectively, these sensors provide complementary visual, attention, and motion cues that make AR devices a strong foundation for intention-aware and context-sensitive interaction~\cite{lv2024ariaeverydayactivitiesdataset}.

Figure~\ref{fig:bg_soc} (c) presents the key components of the AR SoC, including the sensing subsystem, CPU, GPU, DRAM, Image signal processor (ISP) and display engine. In some advanced AR devices, such as the Meta Quest Pro, a neural processing unit (NPU) is also included to support AI workloads. 

\subsection{Embodied Intelligence in AR/VR}
\label{sec:bg:lvu}

A representative application of embodied AI is egocentric video understanding (EVU)~\cite{grauman2024ego, huang2024egoexolearn, mangalam2023egoschema, wang2023holoassist}. Many assistive tasks, such as those illustrated in Figure~\ref{fig:framework}, cannot be resolved from a single image because the relevant evidence unfolds across a sequence of user actions and interactions. More generally, practical embodied intelligence requires preserving sufficient temporal context from perceptual streams so that downstream foundation models, such as the Qwen series~\cite{yang2025qwen3, bai2023qwen}, VideoLlama~\cite{zhang2023videollama}, and others~\cite{jin2023chatunivi,song2023moviechat}, can infer user intent, track task progress, and generate accurate responses. Without EVU, the system loses the continuity needed to provide effective context-aware assistance.


\subsection{Spatial Dynamics of Human Attention}
\label{sec:bg:spatial_temporal}
Human perception is inherently selective, since the brain cannot process the entire visual field with uniformly high fidelity at every moment. 
A simple way to approximate this principle is to crop image patches from the full scene according to the user’s gaze pattern. For example, in Figure~\ref{fig:bg_soc} (d), a user wearing smart glasses walks through a kitchen while looking at a cup on the countertop. Because the user’s attention is centered on the cup, this object is more likely to become the subject of a later query. In contrast, nearby objects, indicated by lighter bounding boxes in Figure~\ref{fig:bg_soc} (d), receive less attention and are therefore less likely to be queried later, as summarized in Figure~\ref{fig:bg_soc} (e). 
However, simple cropping around the gaze location is not the best solution. Although it provides a straightforward way to prioritize attended regions, it can miss important contextual information and cannot fully capture the user’s underlying intent. In Section~\ref{sec:algo_hir}, we describe a machine learning-based method to select and preserve informative regions based on inferred human intention, rather than relying only on intention-based cropping, this enables more accurate and adaptive information selection for accurate EFM operations.

\begin{figure*}
    \centering
    \includegraphics[width=2\columnwidth]{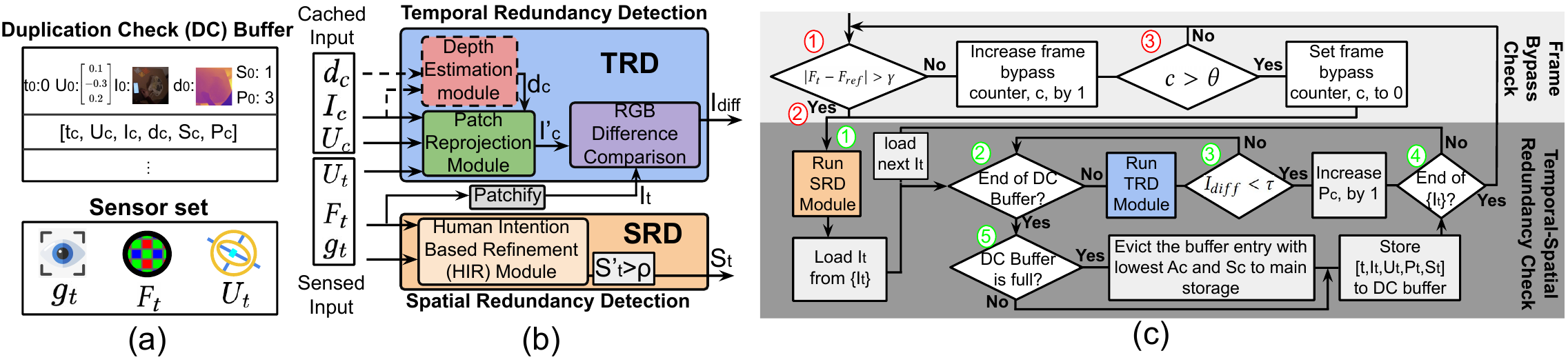}
    \caption{(a) The composition of Duplication Check (DC) Buffer. (b) Temporal redundancy detection module and spatial redundancy detection module. (c) EPIC algorithm. The step numbers for Frame Bypass check (light gray) and spatial-temporal redundancy check (dark gray) are highlighted in~\textcolor{red}{red} and~\textcolor{green}{green}, respectively.}
    \label{fig:algo_main}
\end{figure*}

\subsection{Efficient Egocentric Video Understanding}
\label{sec:bg:efficiency}

In addition to the spatial redundancy discussed in Section~\ref{sec:bg:spatial_temporal}, perceptual video streams also exhibit substantial temporal redundancy. 
Most existing video compression methods operate as offline pipelines that process the full video after recording, such as GenS~\cite{yao2025generative}, Q-Frame~\cite{zhang2025q}, and PruneVid~\cite{huang2025prunevid}. 
In contrast, EPIC is designed for real-time streaming compression, processing and compressing video on the fly as frames are captured. This greatly reduces both video storage and EFM inference latency. Moreover, EPIC leverages AR device-native signals, including user gaze and head pose, to guide compression decisions. Beyond its algorithmic contributions, to the best of our knowledge, EPIC is the~\textbf{first system framework} for efficient video-stream perception in embodied intelligence.

\section{EPIC Algorithm}
\label{sec:EPIC_alg}


\subsection{Geometry-Based Frame Patch Reprojection}

\begin{figure}
    \centering
    \includegraphics[width=0.95\columnwidth]{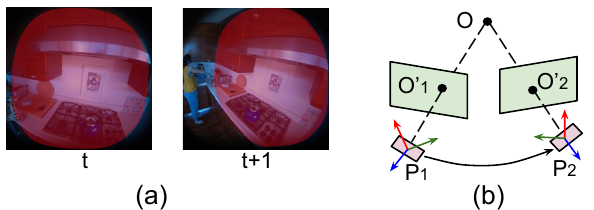}
    \caption{(a) Temporal correlation between consecutive frames. The red areas show the same content observed in different point of views. (b) Perspective reprojection process.}
    \label{fig:algo_proj}
\end{figure}

As shown in Section~\ref{sec:bg:efficiency}, directly computing RGB differences between consecutive frames does not accurately reflect their redundancy. Figure~\ref{fig:algo_proj} (a) shows two consecutive frames $F_{t}$ and $F_{t+1}$ at timesteps $t$ and $t+1$ that remain semantically similar, with most content in frame $t+1$ already present in frame $t$ (highlighted in red), yet their raw RGB difference is still large. This is because user motion changes the viewpoint, introducing substantial pixel-level variation even when the underlying scene content is largely unchanged. To better capture true redundancy, the frames should first be reprojected to compensate for pose variation before computing RGB differences.

Specifically, image formation in a camera can be described by the perspective projection model~\cite{aloimonos1990perspective}. As illustrated in Figure~\ref{fig:algo_proj} (b), a 3D point $O$ in the physical scene is mapped to 2D image points on different image planes, denoted as $O_1'$ and $O_2'$, when observed from two camera positions. Specifically, after the camera moves from $P_1$ to $P_2$, the image of point $O$ shifts from $O_1'$ on the first image plane to $O_2'$ on the second. 
The corresponding 2D coordinates are represented as $o'_{f_1}$ and $o'_{f_2}$, respectively. 
Starting from $o'_{f_1}$, the objective of~\textit{reprojection} is to determine the location of $o'_{f_2}$. To do so, we first recover the 3D position of point $O$ in the coordinate frame of camera position $P_1$. Using the camera intrinsics together with the depth value $d_1$ at $P_1$, the 2D point $O_1'$ is lifted back into 3D space as $o_{p_1}$ through the transformation matrix $T_{cw}(f, d_{1}) \in \mathbb{R}^{4 \times 4}$, where $f$ denotes the focal length. Next, we transform this 3D point into the coordinate frame of camera position $P_2$, obtaining $o_{p_2}$. This step relies on real-time pose information from the IMU, which provides the device translation and orientation. Based on the rotation matrices and translation vectors at the two positions, we derive the transformation matrix $T_{p_1\rightarrow p_2}$ between the two camera frames. Finally, $o_{p_2}$ is projected onto the image plane at $P_2$ using the projection matrix $T_{wc}(f)$, yielding the target image coordinate $o'_{f_2}$. The resulting expression for computing $o'_{f_2}$ from $o'_{f_1}$ is given as:
\begin{equation}
\label{equ:proj}
\small
    [{o'_{f_2}}^T, f, 1]^T =T_{wc}(f)T_{p_1\rightarrow p_2}
    T_{cw}(f,d_1)[{o'_{f_1}}^T, f, 1]^T
\end{equation}
Equation~\ref{equ:proj} can be evaluated for each point in the frame associated with $o'_{f_1}$ to determine its mapped position in $o'_{f_2}$. Because $T_{wc}(f)$, $T_{p_1\rightarrow p_2}$, and $T_{cw}(f,d_1)$ are all represented as $4 \times 4$ matrices, the required computation is relatively inexpensive and parallelizable across all points in the frame. 

Building on this reprojection process, we further eliminate redundant information across frames at a finer granularity by performing redundancy removal at the image-patch level. As shown by the patch reprojection module in Figure~\ref{fig:algo_main} (b), each incoming frame $F_t$ is first divided into a set of fixed-size patches $\{I_t\}$ at timestep $t$. Then, for each buffered patch $I_c$ stored in the~\textit{Duplication Check} (DC) buffer, we reproject it using its buffered pose $U_c$ together with the current pose $U_t$ of $I_t$, producing a reprojected patch $I'_c$. The reprojected buffered patch $I'_c$ and the current patch $I_t$ are then compared through an RGB check, shown as the purple block in Figure~\ref{fig:algo_main} (b), to compute their RGB difference for~\textit{Temporal Redundancy Detection} (TRD). This difference is used to determine whether the current patch should be removed, thereby reducing storage cost and streaming power consumption.


\subsection{Depth Estimation Module}
\label{sec:depth_pred}
Reprojection requires the depth value $d_1$ for each point in the patch. We use the lightweight FastDepth model~\cite{wofk2019fastdepth} to estimate frame depth. To reduce computation, we resize the input image to $64 \times 64$ and interpolate the predicted depth map back to the original resolution. We also quantize the model to 8-bit integers to reduce the memory overhead of the depth prediction module. As shown in Section~\ref{sec:eval:algo}, our evaluation indicates that this design does not affect the performance of EPIC.
As illustrated by the dashed pink block in Figure~\ref{fig:algo_main} (b), depth estimation is performed only once for each buffered image patch $I_{c}$ and the resulting depth map is stored in the DC buffer. The buffered depth $d_c$ is then reused in subsequent TRD operations, further reducing the computational cost of depth estimation.

\subsection{Intention Based Refinement Module}
\label{sec:algo_hir}

As discussed in Section~\ref{sec:bg:spatial_temporal}, human perception is inherently selective, and gaze location provides a strong prior for identifying semantically important regions in egocentric video streams. This signal can help reduce spatial redundancy within each frame $F_t$ by filtering out less important patches $I_t$. However, simple gaze-based subsampling, such as cropping around the gaze location $g_t$, is often insufficient to preserve strong EVU accuracy. To better exploit gaze information, we design a \textit{Human Intention Based Refinement} (HIR) module that uses machine learning to refine image-patch selection.
Specifically, we use a lightweight 3-layer convolutional neural network (CNN) to predict a saliency map for each frame $F_t$. The output is a binary saliency map $S_t$, which indicates patch importance for subsequent operations. This \textit{Spatial Redundancy Detection} (SRD) process is illustrated by the orange block in Figure~\ref{fig:algo_main} (b).

\subsection{Temporal-Spatial Redundancy Check}
\label{sec:algo_red}
Figure~\ref{fig:algo_main} (b) shows the architecture of the TRD module and SRD module. TRD is composed of a depth estimator, a patch reprojection module, and a RGB difference comparison module. The TRD module reprojects buffered patches within the DC buffer to the current viewpoint according to the user’s present pose $U_{t}$.
The structure of DC buffer is illustrated at the top of Figure~\ref{fig:algo_main} (a). Each entry contains six components: the RGB patch $I_c$, its corresponding timestamp $t_c$, pose information $U_c$, depth map $d_c$, the saliency score $S_c$ generated by the HIR module introduced in Section~\ref{sec:algo_hir}, and a popularity score $P_c$. The popularity score records how often patch $I_c$ matches later patches $I_t$; a higher frequency indicates that $I_c$ is more reusable and should be retained for future redundancy checks. Thus, $P_c$ serves as an importance indicator for each patch. Entries in the DC buffer are organized temporally by their timestamps $t_c$.

The workflow of the~\textit{Temporal-Spatial Redundancy Check} (TSRC) is illustrated by the dark gray region in Figure~\ref{fig:algo_main} (c), starting from the top-left corner. At each timestep, the captured video frame $F_t$, together with the associated gaze location $g_t$ and pose information $U_t$, is first passed to the SRD module described in Section~\ref{sec:algo_hir} (\textcolor{green}{step 1}). The SRD module generates a set of image patches $\{I_t\}$ from the current frame $F_t$ along with their corresponding binary saliency map $S_t$.
Next, each patch $I_t$ is checked by the TRD module against the buffered patches $I_c$ stored in the DC buffer, following temporal order from the closest timestep (\textcolor{green}{step 2}). For each comparison, TRD determines whether the current patch is similar to a reprojected cached patch $I_c'$. If the RGB difference $I_{\text{diff}}$ is smaller than a predefined threshold $\tau$, the current patch is considered redundant due to its high contextual similarity to the buffered patch $I_c$, and the popularity score $P_c$ of $I_c$ is incremented by 1 (\textcolor{green}{step 3}). Otherwise, if $I_{\text{diff}} > \tau$, the next entry in the DC buffer is examined. If no matching patch is found after checking all entries, a new entry is inserted into the DC buffer with its popularity score initialized to $P_t = 1$.

\subsection{EPIC Algorithm Summary}
\label{sec:algo_summary}
As discussed in Section~\ref{sec:bg:efficiency}, direct RGB differencing cannot fully capture temporal redundancy across frames, but it provides a lightweight mechanism to filter out trivially redundant inputs. Based on this observation, we further introduce a \textit{Frame Bypass Check}, shown in the light gray region of Figure~\ref{fig:algo_main} (c). This check is applied prior to the TSRC to determine whether the current frame can be skipped entirely. The key intuition is that, during short periods of head stability, consecutive frames exhibit minimal variation and can be safely bypassed without affecting downstream processing.
Specifically, the algorithm first computes the pixel-wise RGB difference between the reference frame $F_{ref}$ and the current frame $F_t$ (\textcolor{red}{Step 1}). If the difference exceeds a threshold $\gamma$, $F_t$ proceeds to the TSRC described in Section~\ref{sec:algo_red} (\textcolor{red}{Step 2}). Otherwise, we use a periodic safeguard to avoid missing subtle but important changes over time. We maintain a counter $c$ that is incremented whenever a frame is bypassed. If $c$ does not exceed a predefined threshold $\theta$, $F_t$ is treated as unchanged and skipped, without invoking the TSRC (\textcolor{red}{Step 3}).

\section{EPIC Hardware System}
\label{sec:hw_design}

\begin{figure*}
    \centering
    \includegraphics[width=1.9\columnwidth]{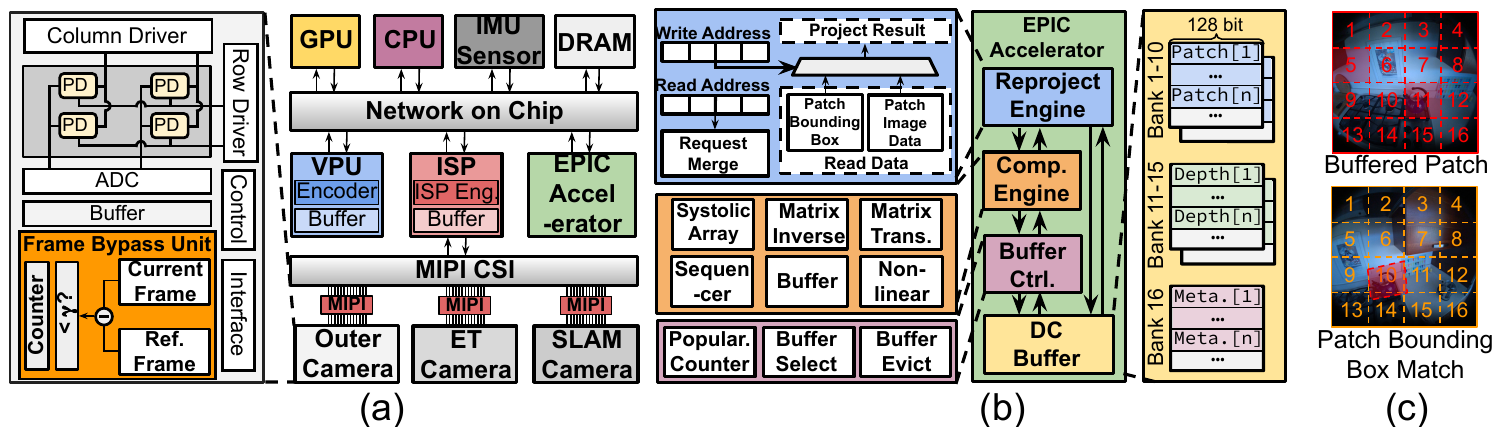}
    \caption{(a) An overview of the EPIC AR SoC. The EPIC accelerator is designed to integrate with the AR SoC. The outer camera is enhanced with a Frame Bypass Unit. (b) EPIC accelerator design. (c) Patch bounding box match. }
    \label{fig:hw_main}
\end{figure*}


In this section, we present the EPIC hardware accelerator for efficient execution of the EPIC algorithm. 
Figure~\ref{fig:hw_main} (a) shows the overall AR SoC architecture. 
The EPIC accelerator is designed as a plug-and-play hardware unit that can be seamlessly integrated into such AR SoCs.



\subsection{EPIC Accelerator}
\label{sec:EPIC_acc}
EPIC accelerator is composed of four major components: the~\textit{reprojection engine},~\textit{computation engine},~\textit{buffer controller}, and~\textit{Duplication Check (DC) buffer}, as shown in Figure~\ref{fig:hw_main} (b). 
\subsubsection{Reprojection Engine Design}
\label{sec:reprojection}
In the EPIC algorithm, buffered patches $I_c$ in the DC buffer are reprojected from their original camera pose $U_c$ to the current pose $U_t$ for duplication checking. Each incoming frame $F_t$ is divided into patches $I_t$, and every patch must be compared with buffered patches iteratively. This exhaustive reprojection process, however, introduces significant computational overhead.
To reduce this overhead, we observe that buffered patches come from different spatial locations within a frame, making full pixel-wise reprojection unnecessary for every patch. Instead, we first reproject only the bounding box of each buffered patch $I_c$ from pose $U_c$ to the current viewpoint, and use the resulting overlap to identify candidate patches $I_t$ for further comparison. Figure~\ref{fig:hw_main} (c) illustrates this process. In the top figure, patches 5 and 11, highlighted in red, are buffered from the previous frame. In the bottom figure, the red region shows the reprojected bounding box of patch 11 in the current frame. Rather than fully reprojecting both patches at the pixel level, only patch 11 is selected for detailed comparison. Since bounding-box reprojection requires much less computation and memory access than full patch reprojection, this mechanism substantially reduces the overhead of redundancy checking in EPIC.

The blue block in Figure~\ref{fig:hw_main} (b) illustrates the reprojection engine, which comprises a write address buffer, a read address buffer, and a request merge unit. 

\subsubsection{Computation Engine and DC Buffer}
\label{sec:computation_engine_dc_buffer}
The computation engine, highlighted in orange in Figure~\ref{fig:hw_main} (b), includes the hardware modules that accelerate geometry-based mask reprojection, depth estimation, and the HIR module. 
The EPIC accelerator includes a dedicated scratchpad for storing DC buffer entries, avoiding contention with the shared system cache and DRAM in the SoC. As described in Section~\ref{sec:algo_red}, each entry stores the RGB patch $I_c$, timestamp $t_c$, pose information $U_c$, depth map $d_c$, saliency score $S_c$ from the HIR module in Section~\ref{sec:algo_hir}, and popularity score $P_c$. The buffer is organized into 16 banks, including 10 for RGB patches, 5 for depth maps, and 1 for metadata. 
The buffer controller updates popularity scores, selects entries, and handles eviction.

\begin{table*}[t]
\centering
\resizebox{\textwidth}{!}{
\begin{tabular}{c|l|cc|cc|cc|cc|cc|cc|cc|cc|cc}
\hline
\multirow{3}{*}{\textbf{Model}} & \multirow{3}{*}{\textbf{\makecell{Base-\\line}}}
  & \multicolumn{6}{c|}{\textbf{EgoEverything}}
  & \multicolumn{6}{c|}{\textbf{HD-Epic}}
  & \multicolumn{6}{c}{\textbf{Nymeria}} \\
\cline{3-20}
& & \multicolumn{2}{c|}{\textbf{Setting 1}} & \multicolumn{2}{c|}{\textbf{Setting 2}} & \multicolumn{2}{c|}{\textbf{Setting 3}} & \multicolumn{2}{c|}{\textbf{Setting 4}} & \multicolumn{2}{c|}{\textbf{Setting 5}} & \multicolumn{2}{c|}{\textbf{Setting 6}} & \multicolumn{2}{c|}{\textbf{Setting 7}} & \multicolumn{2}{c|}{\textbf{Setting 8}} & \multicolumn{2}{c}{\textbf{Setting 9}} \\
\cline{3-20}
& & \textbf{Acc.} & \textbf{Mem.} & \textbf{Acc.} & \textbf{Mem.} & \textbf{Acc.} & \textbf{Mem.}
  & \textbf{Acc.} & \textbf{Mem.} & \textbf{Acc.} & \textbf{Mem.} & \textbf{Acc.} & \textbf{Mem.}
  & \textbf{Acc.} & \textbf{Mem.} & \textbf{Acc.} & \textbf{Mem.} & \textbf{Acc.} & \textbf{Mem.} \\
\hline
\multirow{5}{*}{\rotatebox{90}{\makecell{Qwen2.5\\-VL-7B}}}
  & FV   & 61.4\% & 19.9$\times$ & 61.4\% & 43.3$\times$ & 61.4\% & 107.5$\times$ & 53.3\% & 4.44$\times$ & 53.3\% & 5.27$\times$ & 53.3\% & 7.39$\times$ & 65.7\% & 20.2$\times$ & 65.7\% & 40.2$\times$ & 65.7\% & 103.1$\times$ \\
& SD   & 45.4\% & 1.04$\times$ & 42.4\% & 1.10$\times$ & 43.2\% & 1.10$\times$ & 43.4\% & 1.12$\times$ & 38.9\% & 1.02$\times$ & 34.5\% & 1.11$\times$ & 53.4\% & 1.03$\times$ & 51.5\% & 1.04$\times$ & 46.9\% & 1.06$\times$ \\
& TD   & 54.2\% & 1.02$\times$ & 51.8\% & 1.04$\times$ & 48.2\% & 1.05$\times$ & 47.6\% & 1.03$\times$ & 45.3\% & 1.01$\times$ & 41.3\% & 1.04$\times$ & 61.2\% & 1.03$\times$ & 58.3\% & 1.02$\times$ & 56.6\% & 1.00$\times$ \\
& GC   & 53.3\% & 1.06$\times$ & 52.7\% & 1.14$\times$ & 41.5\% & 1.11$\times$ & 45.6\% & 1.08$\times$ & 41.6\% & 1.07$\times$ & 35.2\% & 1.19$\times$ & 51.6\% & 1.03$\times$ & 50.2\% & 1.03$\times$ & 42.5\% & 1.06$\times$ \\
\rowcolor{blue!10}\cellcolor{white}& \textbf{EPIC} & 59.9\% & 1$\times$ & 57.2\% & 1$\times$ & 56.1\% & 1$\times$ & 51.3\% & 1$\times$ & 50.8\% & 1$\times$ & 48.2\% & 1$\times$ & 65.8\% & 1$\times$ & 63.8\% & 1$\times$ & 62.5\% & 1$\times$ \\
\hline
\end{tabular}
}
\caption{EVU accuracy (Acc.) and normalized memory footprint (Mem.) results. Memory footprint is normalized to EPIC ($1\times$).}
\label{tab:main_results}
\end{table*}

\subsection{In-sensor Frame Bypass Unit}
\label{sec:sensor_design}
With growing interest in in-sensor and near-sensor computing, image sensors have become a promising platform for accelerating AR/VR workloads~\cite{an2020ultra, tsai2025400, sun2024estimating, liu2022augmented}. In EPIC, the Frame Bypass Check is implemented inside the image sensor as a dedicated~\textit{Frame Bypass Unit}, which computes the pixel-wise difference between a stored reference frame $F_{ref}$ and the incoming frame $F_t$. 

The Frame Bypass Unit is depicted in the orange block in Figure~\ref{fig:hw_main} (a). The image sensor stores a reference frame $F_{ref}$ in an on-chip buffer. As pixels from $F_t$ are digitized by the ADC, each is immediately compared with its corresponding pixel in $F_{ref}$ using subtraction and thresholding. If the frame-level difference remains below $\gamma$, the frame is treated as visually redundant. To avoid excessive skipping, a counter enforces a minimum frame preservation rate, guaranteeing that at least one frame is sent to the SoC within a bounded interval, 
as described in Section~\ref{sec:algo_summary}. 

\section{Accuracy Evaluation}
\label{sec:eval:algo}

In this section, we evaluate the accuracy performance of EPIC algorithm on three EVU datasets: EgoEverything~\cite{tang2026egoeverythingbenchmarkhumanbehavior}, HD-Epic~\cite{Perrett_2025_CVPR}, and Nymeria~\cite{ma2024nymeria}. All three datasets contain multiple-choice questions derived from video clips sampled at 10 FPS. EgoEverything clips are capped at 3 minutes, HD-Epic clips are trimmed using the start and end timestamps of each question, and Nymeria video clips have an average length of 10 minutes and in some cases exceeding 30 minutes. These durations already cover most EVU use cases in daily life, and EPIC can naturally extend to longer streams as such data become available. The CNN in HIR module is fine-tuned on 1000 questions from each dataset using splits disjoint from the test set. We evaluate the accuracy on one EFMs: Qwen2.5-VL-7B~\cite{bai2025qwen25vltechnicalreport}.
We compare against four baseline video compression algorithms. Full Video (FV) retains the video at the original FPS and spatial resolution. Spatial Downsample (SD) maintains the original FPS while uniformly downsampling frames spatially to match a target memory budget. Temporal Downsample (TD) preserves the original frame resolution while uniformly skipping frames to achieve the target memory size. Gaze Crop (GC) crops a square region centered at the gaze point from each frame, retaining only the cropped region to meet the target memory budget. 

\subsection{Egocentric Video Understanding Accuracy}
\label{sec:vqa_acc}

As shown in Table~\ref{tab:main_results}, EPIC consistently achieves higher EVU accuracy than all baseline methods while maintaining the lowest memory footprint across all datasets and models. Compared with FV, EPIC reduces memory footprint by $97.6\%$, $82.2\%$, and $97.5\%$ on EgoEverything, HD-Epic, and Nymeria, respectively, while incurring an average accuracy drop of only $3.0\%$, $2.4\%$, and $3.2\%$. Compared with SD, TD, and GC baselines at equivalent memory budgets, EPIC achieves on average $12.9\%$, $5.1\%$, and $12.1\%$ higher accuracy across all datasets and models. This consistent improvement demonstrates that EPIC algorithm effectively retains task-relevant visual information, avoiding the uniform quality degradation introduced by SD, TD and GC.

\section{Hardware Performance Evaluation}
\label{sec:hardware-eval}
\begin{figure}
    \centering
    \includegraphics[width=1\columnwidth]{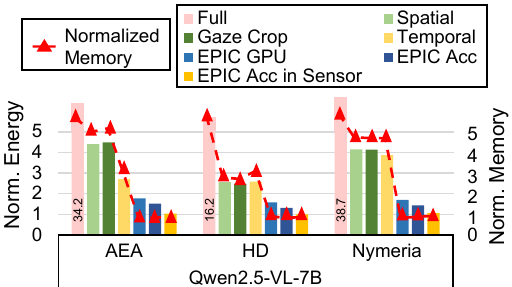}
    \caption{Energy consumption (shown in bars) and memory footprint (red lines) evaluation across different methods.}
    \label{fig:hw_eval_main}
\end{figure}
In this section, we evaluate the hardware performance of the EPIC accelerator. As shown in Figure~\ref{fig:hw_main} (b), the accelerator consists of a $16\times16$ 2D systolic array, non-linear units, a reprojection engine, a buffer controller, and supporting interfaces, all implemented in SystemVerilog at 1~GHz. We synthesize the design using Synopsys Design Compiler~\cite{Baliga2019Synopsys} and evaluate area, timing, and power through cycle-accurate RTL simulation in 45nm CMOS technology~\cite{nangate}. A 4~MB on-chip SRAM serves as the DC buffer, and an additional 768~KB SRAM stores model weights and activations. 

\subsection{System Evaluation Result}
\label{sec:eval-hw-result}
To evaluate end-to-end system performance, we compare against several baseline configurations. \textbf{Full Video System (FVS)} captures all frames at the original frame rate and resolution, transmits the full stream over MIPI, compresses it with H.264 on the VPU, and stores the result. \textbf{Gaze Crop System (GCS)}, \textbf{Spatial Downsample System (SDS)}, and \textbf{Temporal Downsample System (TDS)} are configured to match EPIC’s EVU accuracy using the settings identified in Section~\ref{sec:vqa_acc}. 
\textbf{EPIC+GPU} runs the full EPIC algorithm on the Qualcomm Adreno GPU of the Qualcomm Open-Q 865 board, without the EPIC accelerator or in-sensor processing. \textbf{EPIC+Acc} instead offloads the full algorithm to the dedicated EPIC accelerator.
Finally, \textbf{EPIC+Acc+In-Sensor} further enables the Frame Bypass Unit inside the image sensor.

The hardware performance results are shown in Figure~\ref{fig:hw_eval_main}. 
\textbf{EPIC+Acc+In-Sensor} achieves the lowest energy consumption and memory footprint. Compared with \textbf{FVS}, it reduces energy by $24.3\times$ and memory footprint by $27.5\times$ on average while maintaining comparable EVU accuracy, as shown in Table~\ref{tab:main_results}. Under similar EVU accuracy, \textbf{EPIC+Acc+In-Sensor} further reduces energy by $2.40\times$, $3.09\times$, and $3.08\times$ over TDS, SDS, and GCS, respectively, and reduces memory footprint by $3.28\times$, $4.03\times$, and $4.00\times$, demonstrating EPIC’s advantage.

\section{Conclusion}
\label{sec:conclusion}
We present EPIC, an efficient egocentric perception system for embodied intelligence on smart AR glasses. By jointly optimizing algorithms, hardware, and in-sensor processing, EPIC removes redundant video content and significantly reduces energy and memory costs while preserving accuracy. 

\newpage

\bibliographystyle{ACM-Reference-Format}
\bibliography{refs}
\appendix

\end{document}